\pdfoutput=1
\documentclass[11pt,a4paper]{article}
\usepackage[T1]{fontenc}
\usepackage[utf8]{inputenc}
\usepackage{lmodern}
\usepackage[margin=22mm]{geometry}
\usepackage{microtype}
\usepackage{amsmath,amssymb}
\usepackage{booktabs}
\usepackage{tabularx}
\usepackage{array}
\usepackage{graphicx}
\usepackage{float}
\usepackage{caption}
\usepackage{subcaption}
\usepackage{enumitem}
\usepackage{xcolor}
\usepackage{tikz}
\usepackage{pgfplots}
\usetikzlibrary{arrows.meta,positioning,calc,fit,backgrounds,shapes.geometric,patterns,decorations.pathreplacing}
\usepgfplotslibrary{groupplots}
\pgfplotsset{compat=1.16}
\usepackage[numbers,sort&compress]{natbib}
\usepackage{url}
\usepackage[colorlinks=true,linkcolor=black,citecolor=black,urlcolor=black]{hyperref}
\newif\ifsubmission \submissionfalse
\ifsubmission\usepackage[mathlines]{lineno}\fi
\setlist{itemsep=2pt,topsep=3pt}
\newcolumntype{Y}{>{\raggedright\arraybackslash}X}
\newcommand{\code}[1]{\texttt{#1}}
\newcommand{\woma}{\textsf{woma}}

\usepackage{afterpage}
\usepackage[section]{placeins}
\usepackage{fancyhdr}
\fancypagestyle{plain}{\fancyhf{}\fancyhead[R]{\small\thepage}}

\definecolor{cTeal}{gray}{0.15}
\definecolor{cBlue}{gray}{0.30}
\definecolor{cGreen}{gray}{0.05}
\definecolor{cGrey}{gray}{0.45}
\definecolor{cRed}{gray}{0.20}
\definecolor{cAmber}{gray}{0.55}
\definecolor{cLight}{gray}{0.94}
\tikzset{
  blk/.style   = {draw=cGrey!75, fill=white, rounded corners=1.5pt,
                  font=\scriptsize, align=center, inner sep=2.5pt, minimum height=6mm},
  stg/.style   = {draw=cGrey!75, fill=cLight, rounded corners=1.5pt,
                  font=\scriptsize, align=center, inner sep=2.5pt, minimum height=6mm},
  stgh/.style  = {stg, draw=cTeal!80, fill=cTeal!10},
  mod/.style   = {draw=cGrey!75, fill=cLight, rounded corners=2pt,
                  font=\scriptsize, align=center, inner sep=3pt, minimum height=8mm},
  sum/.style   = {draw=cGrey!75, circle, inner sep=0pt, minimum size=4.4mm,
                  font=\scriptsize},
  tapn/.style  = {draw=cGreen!80, fill=cGreen!10, rounded corners=1pt,
                  font=\tiny, inner sep=1.5pt},
  flow/.style  = {-{Stealth[length=2mm]}, draw=cGrey, line width=0.5pt},
  idf/.style   = {-{Stealth[length=2mm]}, draw=cGreen, line width=1.1pt,
                  rounded corners=2pt},
  killf/.style = {-{Stealth[length=2mm]}, draw=cRed, line width=0.9pt,
                  dashed},
  gate/.style  = {draw=cAmber, fill=cAmber!10, diamond, aspect=2.2,
                  font=\scriptsize, align=center, inner sep=1.5pt},
  pstep/.style = {draw=cGrey!75, fill=white, rounded corners=2pt,
                  font=\scriptsize, align=center, inner sep=3pt},
  killb/.style = {draw=cRed, dashed, fill=cRed!8, rounded corners=2pt,
                  font=\scriptsize, align=center, inner sep=3pt},
  lane/.style  = {draw=cGrey!50, fill=cLight!60, rounded corners=3pt,
                  inner sep=4pt},
  gbar/.style  = {draw=cTeal!70, fill=cTeal!30, rounded corners=1pt},
  gbarx/.style = {draw=cAmber!90, pattern=dots, pattern color=cAmber!90,
                  rounded corners=1pt},
  gbarr/.style = {draw=cRed!70, pattern=north east lines,
                  pattern color=cRed!70, rounded corners=1pt},
  done/.style  = {circle, fill=cGreen, inner sep=1.6pt},
  todo/.style  = {circle, draw=cGrey, inner sep=1.6pt, fill=white},
  lbl/.style   = {font=\scriptsize, text=cGrey},
}

\title{\bfseries \woma: a real-time foundation model\\and its fine-tuned models for endoscopy}

\author{%
\begin{tabular}{c@{\hspace{14mm}}c}
Thang Tran\thanks{Corresponding author: \texttt{thang.tran@cloudkites.com}} & Lan Dang \\
\small CloudKites AI Lab & \small Monash Business School, Monash University \\
\small New South Wales, Australia & \small Victoria, Australia \\
\small\texttt{thang.tran@cloudkites.com} & \small\texttt{LanHong.Dang@monash.edu}
\end{tabular}}

\date{September 2026}

\begin{document}
\ifsubmission\linenumbers\fi
\maketitle

\begin{abstract}
\woma{} is a real-time foundation model for gastrointestinal endoscopy: a network
trained without labels on about a million endoscopy frames, from which task models are
fine-tuned. We contribute a systematic design for production. Requirements and pass marks were fixed
before any run, eight candidates screened under pre-registered rules, self-supervised training
taken to a stopping rule, then fine-tuning and deployment optimisation, all on one self-contained
library, numbat. We also contribute \woma{} itself with two fine-tuned models, every outcome
reported met or missed. Our colonoscopy model finds and outlines polyps, names which colon
segment is in view, suggests polyp type and grades bowel preparation. Our gastroscopy model names
a station out of 22 protocol sites, flags and outlines lesions, and names one of seven findings.
Every number was read on data never seen in training, and shipped weights were chosen on that record.
In colonoscopy, 96\,\% of polyps in a six-hospital PolypGen set are found at precision
$\geq$0.85, and 19 of 19 polyps across fifteen full REAL-Colon videos at 1.6 false alarms per
procedure. In gastroscopy, landmark region is named correctly on 92\,\% of frames from unseen
patients, and 37 of 39 held-out neoplasia frames are flagged at specificity 0.91. On one
workstation GPU every task runs over 1080p video at about 100 frames per second, faster than
PyTorch, ONNX Runtime and TensorRT in all four precision regimes tested. TensorRT comes closest:
one pass of our foundation model takes it 3 to 27\,\% longer than ours, and we deliver 6 to
31\,\% more frames per second from frame to results. A second build links no vendor library at
all --- our own kernels over Vulkan --- so a site deploys two files and needs no toolkit, no
cuDNN and no framework; in f32 it beats the CUDA build on the same card.
\end{abstract}

\vspace{2mm}
\noindent\textbf{Keywords:} foundation model; self-supervised learning; real-time inference; colonoscopy; gastroscopy

\section{Introduction}
\label{sec:intro}

Computer-aided polyp detection has crossed from research into practice: a randomised
trial showed a real-time detector raising the adenoma detection rate
\citep{repici2020cade}, a meta-analysis confirmed the effect \citep{hassan2021meta}, and
a first device is sold with its design and validation described in print
\citep{cherubini2023gigenius}. In the stomach, a system that watches for unexamined
\emph{blind spots} cut the blind-spot rate from 22\,\% to 6\,\% in a randomised trial
\citep{wu2019wisense}, and a convolutional network found gastric cancer with a
per-lesion sensitivity of 92\,\% \citep{hirasawa2018gastric}. What such reports rarely
give is the whole path: how the network at the centre was chosen, what trained each head,
what each number was read on, how the shipped weights were picked, and how fast the
system runs when measured rather than estimated. This paper gives that path for one
system, and it contributes two things.

\begin{enumerate}
\item \textbf{A systematic design for production, from the first day.} Requirements and pass
 marks were written before anything ran; eight candidate architectures were screened under
 pre-registered rules; data pipeline, trainer, label-free training to a stopping rule,
 fine-tuning of task models and deployment optimisation all ran on one self-contained library,
 numbat \citep{tran2026numbat,tran2026crossstack}; and what came out is benchmarked against
 PyTorch, ONNX Runtime and TensorRT on one machine, for our foundation model and for both
 fine-tuned models end to end.
\item \textbf{A real-time foundation model and its fine-tuned models for endoscopy, with
 transparent outcomes.} \woma{}, with colon and upper-GI models built on it, is reported
 against every pass mark met or missed, on data training never saw, alongside rival
 architectures fine-tuned on identical recipes.
\end{enumerate}

Section~\ref{sec:methods} covers requirements, data, design experiment, fine-tuning recipes
and evaluation protocol. Section~\ref{sec:results} reports screening outcome, held-out results
and speed. Section~\ref{sec:discussion} weighs what both contributions are worth, why held-out
evaluation matters, and what locally collected data would add. Every number here is
retrospective, on public data.

\section{Methods}
\label{sec:methods}

\subsection{Requirements, targets and criteria}
\label{sec:problem}

\begin{figure}[t]
\centering
\resizebox{\textwidth}{!}{%
\begin{tikzpicture}[x=1cm,y=1cm]
\node[mod, minimum width=17mm] (vid) at (0,0) {60 fps\\video};
\node[mod, minimum width=19mm, right=6mm of vid] (pre) {preprocess\\{\scriptsize crop, letterbox}};
\node[mod, minimum width=24mm, right=6mm of pre, fill=cTeal!12, draw=cTeal!70] (bb) {\textbf{foundation}\\\textbf{model} (woma)};
\node[mod, minimum width=19mm, right=6mm of bb] (neck) {feature\\pyramid neck};
\node[mod, minimum width=21mm, above right=1.5mm and 6mm of neck] (det) {detection head};
\node[mod, minimum width=21mm, right=6mm of neck] (seg) {segmentation\\{\scriptsize region-gated}};
\node[mod, minimum width=21mm, below right=1.5mm and 6mm of neck] (sta) {station /\\quality heads};
\node[mod, minimum width=17mm, right=32mm of neck] (tmp) {temporal\\filter};
\node[mod, minimum width=15mm, right=6mm of tmp] (ovl) {overlay};
\draw[flow] (vid)--(pre); \draw[flow] (pre)--(bb); \draw[flow] (bb)--(neck);
\draw[flow] (neck)--(det); \draw[flow] (neck)--(seg);
\draw[flow] (neck.south east)--(sta.west);
\draw[flow] (det.east) -- (tmp.north west);
\draw[flow] (seg.east) -- (tmp.west);
\draw[flow] (sta.east) -- (tmp.south west);
\draw[flow] (tmp)--(ovl);
\node[tapn, below=1mm of bb] {$P_2\,P_3\,P_4\,P_5$ + pooled};
\node[lbl, below=16mm of neck, xshift=-4mm, align=center, font=\scriptsize]
 {frame budget 16.7 ms: model+neck $\leq 8$ \quad heads $\approx 5$
  \quad temporal $\approx 2$ \quad margin $\approx 1.7$};
\end{tikzpicture}}
\caption{Where the foundation model sits. Every head reads its features, and the time budget of one
video frame is split so that the foundation model and neck may spend at most about half of it.}
\label{fig:where}
\end{figure}
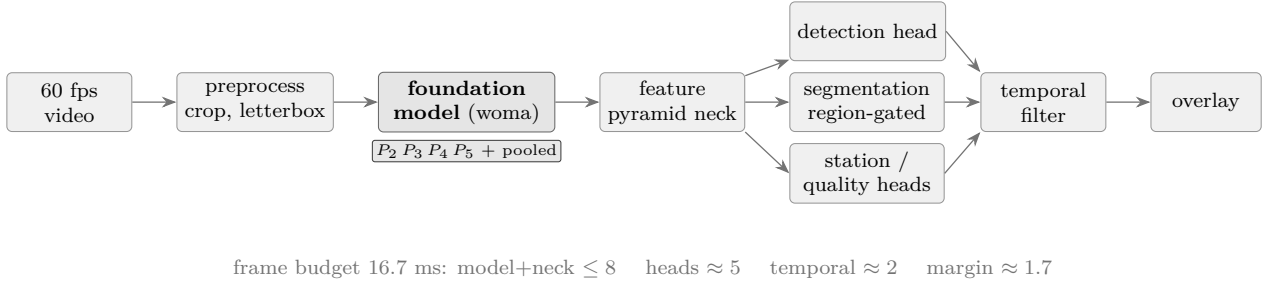

The \emph{foundation model} is the part of the network that turns a video frame into
features; every task head reads those features (Figure~\ref{fig:where}). At 60 frames per
second a frame lasts 16.7 milliseconds (ms), and the foundation model and its
feature-pyramid neck may spend at most about half of that on a 640$\times$640 pixel input,
on one workstation graphics processing unit (GPU). Sixty is the demanding end of what endoscopy
processors emit, and a stack that keeps up with it keeps up with the 25, 30 and 50 frames per
second that others produce, so one budget serves every site. It is a requirement on deployment
rather than a property of our material --- nothing here was recorded at 60, and the two clips the
delivered program is timed on run at 15 and 30 frames per second (Table~\ref{tab:speed}) --- so
speed is measured against a fixed input under the protocol of Section~\ref{sec:evalprotocol},
never inferred from the recordings the accuracy numbers come from. Beyond speed, it must serve four kinds
of head at once (finding lesions, outlining them, naming the anatomical site and grading
quality), it must do so in hospitals it never saw, and it must stay quiet on clean mucosa,
because an alert that fires on nothing is what makes clinicians switch a system off. Pass marks came from published anchors, fixed before any run (Table~\ref{tab:targets}).

\newcommand{\dsbox}[1]{\parbox[c][9.5mm][c]{36mm}{\raggedright\scriptsize #1}}
\newcommand{\rolebox}[1]{\parbox[c][17.5mm][c]{42mm}{\centering\scriptsize #1}}
\begin{figure}[t]
\centering
\resizebox{0.9\textwidth}{!}{%
\begin{tikzpicture}[x=1cm,y=1cm,
  ds/.style={draw=cGrey!75, fill=white, rounded corners=1.5pt, inner sep=2.5pt},
  role/.style={draw=cTeal!80, fill=cTeal!8, rounded corners=2pt, inner sep=2.5pt},
]
\node[ds] (gn)  at (0, 0.00) {\dsbox{\textbf{GastroNet-5M} --- 1.0M held of 4.8M, 8 centres}};
\node[ds] (own) at (0,-1.36) {\dsbox{\textbf{Own corpus} --- $\sim$0.5M frames, growing}};
\node[ds] (hk)  at (0,-2.72) {\dsbox{\textbf{HyperKvasir} --- 99k unlabelled + 10.6k labelled}};
\node[ds] (il)  at (0,-4.08) {\dsbox{\textbf{Internal labels} --- detection; EGD stations}};
\node[ds] (gh)  at (0,-5.44) {\dsbox{\textbf{GastroHUN} --- 8.8k images, 22 stations}};
\node[ds] (kv)  at (0,-6.80) {\dsbox{\textbf{Kvasir-SEG} --- 1k segmentation masks}};
\node[ds] (ss)  at (0,-8.16) {\dsbox{\textbf{SUN-SEG} --- 158k video frames}};
\node[ds] (pg)  at (0,-9.52) {\dsbox{\textbf{PolypGen} --- 8k+ images, 6 centres}};
\node[ds] (rc)  at (0,-10.88) {\dsbox{\textbf{REAL-Colon} --- 60 full videos, 2.7M frames}};
\node[role] (pool)  at (8.6,-0.60) {\rolebox{\textbf{pretraining pool}\\$\sim$1.5M frames held --- self-supervision and distillation targets}};
\node[role] (probe) at (8.6,-3.05) {\rolebox{\textbf{screening probes}\\frozen detection $+$ frozen stations, at fixed intervals}};
\node[role] (ft)    at (8.6,-5.50) {\rolebox{\textbf{fine-tuning}\\detection, segmentation and station heads at $640^2$}};
\node[role, draw=cGrey!70, fill=cLight] (tmp) at (8.6,-7.95) {\rolebox{\textbf{temporal evaluation}\\video segmentation consistency}};
\node[role, draw=cRed!70, fill=cRed!6]  (ho)  at (8.6,-10.40) {\rolebox{\textbf{held-out evaluation --- never trained on}\\unseen-centre accuracy; false alarms per procedure}};
\draw[flow] (gn.east) -- (pool.west);
\draw[flow] (own.east) -- (pool.west);
\draw[flow] (hk.east) -- (pool.west);
\draw[flow] (hk.east) -- (probe.west);
\draw[flow] (il.east) -- (probe.west);
\draw[flow] (il.east) -- (ft.west);
\draw[flow] (gh.east) -- (probe.west);
\draw[flow] (gh.east) -- (ft.west);
\draw[flow] (kv.east) -- (ft.west);
\draw[flow] (ss.east) -- (tmp.west);
\draw[flow, draw=cRed!70] (pg.east) -- (ho.west);
\draw[flow, draw=cRed!70] (rc.east) -- (ho.west);
\end{tikzpicture}}
\caption{Data pool: every dataset, its size, and the single role it plays. The bottom lane
is walled off --- PolypGen and REAL-Colon feed evaluation only, so the unseen-centre and
false-alarm numbers mean what they say. Sources: GastroNet-5M \citep{boers2024gastronet},
HyperKvasir \citep{borgli2020hyperkvasir}, GastroHUN \citep{bravo2025gastrohun}, Kvasir-SEG
\citep{jha2020kvasirseg}, SUN-SEG \citep{misawa2021sun,ji2022vps}, PolypGen
\citep{ali2023polypgen}, REAL-Colon \citep{biffi2024realcolon}.}
\label{fig:dataflow}
\end{figure}
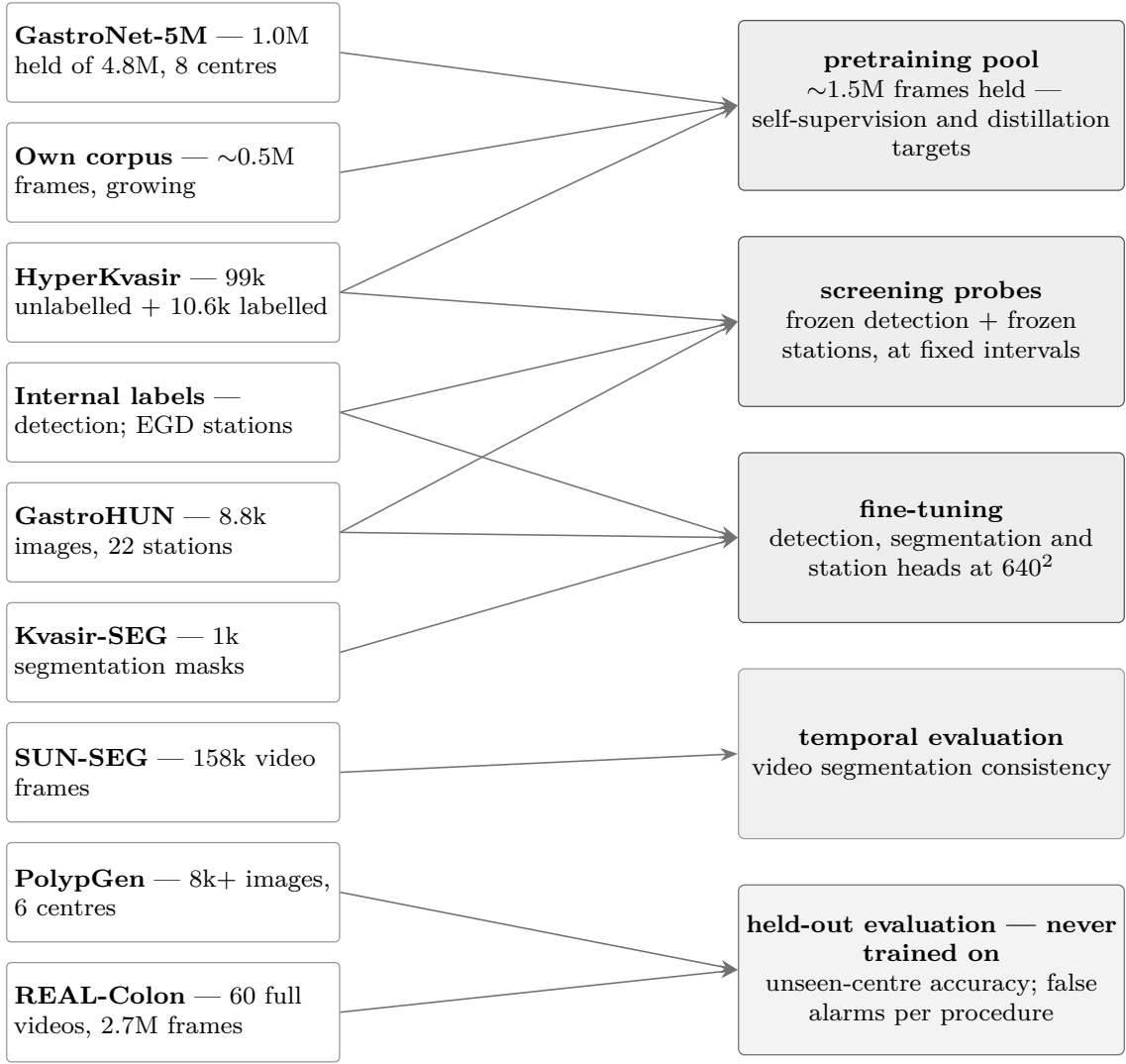
\begin{figure}[p]
\centering
\resizebox{\textwidth}{!}{%
\begin{tikzpicture}[x=1cm, y=1cm, every node/.append style={font=\small}]
\node[pstep, text width=70mm, draw=cGreen!70] (ph0) at (0,0)
 {\textbf{Design study}\\the problem, the criteria, a review of the field,
  eight candidates, the plan};
\node[pstep, text width=70mm] (ph1) at (0,-1.9)
 {\textbf{Stage 0: measure}\\latency of every deployable form $\cdot$ training
  throughput $\cdot$ probe calibration};
\node[pstep, text width=70mm] (ph2) at (0,-3.6)
 {\textbf{Builds}\\eight candidates in the training harness, the same probes on all};
\node[pstep, text width=70mm] (ph3) at (0,-5.5)
 {\textbf{Stage 1: screening}\\aspects I to IV on all eight: equal-clock twelve-hour
  runs, matched-images re-probe, route-optimal wave, adaptability fine-tune};
\node[gate, text width=30mm] (g1) at (0,-7.55) {ranked ladder:\\rungs 1--2 advance};
\node[pstep, text width=70mm] (ph4) at (0,-9.45)
 {\textbf{Stage 2: confirmation}\\72-hour runs, both finalists in parallel,
  plus a fine-tune smoke test};
\node[gate, text width=30mm] (g2) at (0,-11.5) {winner by\\pre-registered rank};
\node[pstep, text width=70mm] (ph5) at (0,-13.4)
 {\textbf{Production foundation model}\\the winner's route, run to its stopping rule on the full pool};
\node[pstep, text width=70mm] (ph6) at (0,-15.1)
 {\textbf{Multi-task fine-tune}\\detection, outline, station and findings heads on the labelled sets};
\node[pstep, text width=70mm] (ph7) at (0,-16.8)
 {\textbf{Packaging}\\one model file per model; latency verified against the budget};
\node[pstep, text width=70mm] (ph8) at (0,-18.6)
 {\textbf{Evaluation on held-out data}\\PolypGen unseen centres $\cdot$ REAL-Colon false alarms per
  procedure $\cdot$ GastroHUN stations $\cdot$ Kvasir-SEG};
\node[pstep, text width=70mm, draw=cTeal!80] (ph9) at (0,-20.4)
 {\textbf{Report}\\this paper};
\draw[flow] (ph0)--(ph1); \draw[flow] (ph1)--(ph2);
\draw[flow] (ph2)--(ph3); \draw[flow] (ph3)--(g1);
\draw[flow] (g1)--(ph4); \draw[flow] (ph4)--(g2);
\draw[flow] (g2)--(ph5); \draw[flow] (ph5)--(ph6);
\draw[flow] (ph6)--(ph7); \draw[flow] (ph7)--(ph8);
\draw[flow] (ph8)--(ph9);
\node[killb, text width=36mm] (k1) at (5.7,-7.55)
 {a run below the floor, a rank collapse, or a station probe at chance: killed};
\draw[killf] (g1)--(k1);
\node[killb, text width=36mm] (k2) at (5.7,-11.5)
 {all fail: fall back to the strongest measured deploy stack, trained with labels only};
\draw[killf] (g2)--(k2);
\node[pstep, text width=36mm, draw=cTeal!70] (cands) at (-5.7,-5.5)
 {\textbf{the eight candidates}\\1 R34 $\cdot$ 2 RVGG $\cdot$ 3 CNXV2\\
  4 FMS $\cdot$ 5 CSPR $\cdot$ 6 CSPR-A\\7 RVT $\cdot$ 8 CSPR-S};
\draw[flow, draw=cTeal] (cands)--(ph3);
\end{tikzpicture}}
\caption{Whole project from the design study to the evaluated models, with its two
gates. Every exit is a decision that was written down before anything ran; the two
``killed'' branches are outcomes we were prepared to take.}
\label{fig:master}
\end{figure}
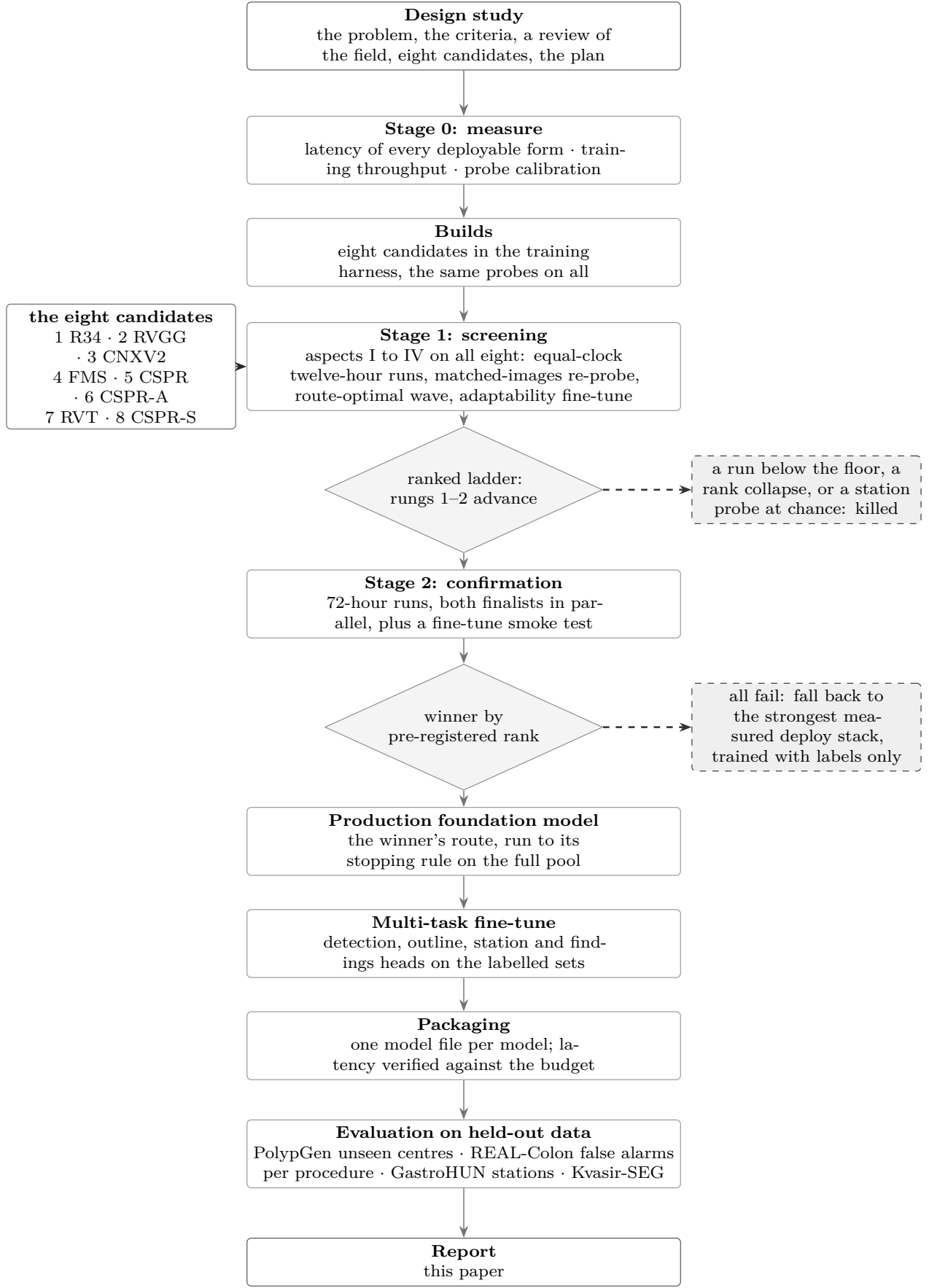

\begin{table}[t]
\centering\footnotesize
\caption{Pass marks, fixed before any run, and the published anchor each was set from. One row
per mark; the corpora they are read on are named in Section~\ref{sec:evalprotocol}.}
\label{tab:targets}
\setlength{\tabcolsep}{5pt}
\begin{tabular}{p{7.0cm} r p{5.8cm}}
\toprule
instrument & mark & anchor \\
\midrule
\multicolumn{3}{@{}l}{\emph{colonoscopy}} \\
PolypGen sensitivity, precision $\geq$0.85 & $\geq$0.90 & cross-centre transfer \citep{ali2023polypgen} \\
REAL-Colon per-polyp sensitivity & $\geq$0.93 & per-procedure reporting \citep{biffi2024realcolon} \\
outline Dice, three other-hospital sets & $\geq$0.82 & PraNet protocol \citep{fan2020pranet} \\
colon segment macro recall, ten segments & $\geq$0.70 & single-frame best \citep{saito2021anatomical} \\
colon segment accuracy, six merged classes & 0.666 & the same, at parity \citep{saito2021anatomical} \\
polyp type macro recall & $\geq$0.80 & reported, no published parity \\
bowel preparation adequacy $F_1$ & $\geq$0.90 & Boston scale \citep{lai2009bbps} \\
\midrule
\multicolumn{3}{@{}l}{\emph{gastroscopy}} \\
station landmark-region accuracy & $\geq$0.92 & 0.887, the dataset paper \citep{bravo2025gastrohun} \\
station 23-code macro recall & $\geq$0.85 & the same, reported \\
lesion gate sensitivity, specificity $\geq$0.90 & $\geq$0.93 & 0.922 per image \citep{hirasawa2018gastric} \\
findings macro recall, validation & $\geq$0.85 & 0.916--0.939, FastUGI-Net \citep{chan2025fastugi} \\
findings accuracy, UGIAD coarse-5 & $\geq$0.90 & the same \\
findings accuracy, GastroVision half & $\geq$0.85 & the same \\
\midrule
\multicolumn{3}{@{}l}{\emph{speed, one workstation GPU}} \\
whole stack, per frame & $\leq$16.7\,ms & $\approx$10\,ms, commercial \citep{cherubini2023gigenius} \\
foundation model and neck, per frame & $\leq$8\,ms & the same \\
\midrule
\multicolumn{3}{@{}p{0.97\textwidth}@{}}{\scriptsize The detection marks are read on
PolypGen's six hospitals \citep{ali2023polypgen,ali2024endocv2021} and on REAL-Colon study 4 at
no more than two sustained false alarms per procedure \citep{biffi2024realcolon,holzwanger2021fp};
the outline mark is the mean over CVC-ColonDB \citep{bernal2012colondb}, ETIS
\citep{silva2014etis} and CVC-300 \citep{vazquez2017endoscene}; the lesion gate is 39 neoplasia
frames against 1{,}347 clean frames from another hospital. A mark with no published parity is
still fixed in advance and reported either way.} \\
\bottomrule
\end{tabular}
\end{table}

\subsection{Data}
\label{sec:data}

Figure~\ref{fig:dataflow} shows the data: about 1.5 million unlabelled frames for
pretraining, labelled sets for the heads, and two sets walled off from training from the
first day. PolypGen \citep{ali2023polypgen} holds polyp images from six hospitals and was
built to test the unseen-hospital question; REAL-Colon \citep{biffi2024realcolon} holds
sixty full-length colonoscopy videos with histology for every polyp. Both are read only.
Everything else is split by video or by patient, because consecutive endoscopy frames are
near-identical: a split by frame once inflated our detection score from 0.67 to a
fictitious 0.91. The colon and upper-GI training sets, their exclusions and their routing
rules are given in Figures~\ref{fig:wfcolon} and \ref{fig:wfupper}.

\subsection{Designing the foundation model}
\label{sec:design}

\paragraph{Learning without labels, and three routes.} A network can learn from unlabelled
frames by solving a task that needs no labels, such as filling in a masked part of the
image or agreeing with itself across two views of the same image.
Figure~\ref{fig:objectives} sorts objectives by three properties that matter here:
whether the recipe works on a \emph{convolutional} network, the kind that fits a 16.7\,ms
budget on a small GPU; whether it teaches every location of the feature map, which
detection and outlining need; and whether it guards against \emph{collapse}, the failure
in which every feature drifts to the same value. Lesions are recognised by texture, which
favours objectives that must rebuild texture over objectives that learn to ignore it.
There are three routes to a pretrained network (Figure~\ref{fig:routes}): from a random
start on the unlabelled pool, from public weights continued on the pool, or by distilling a
large foundation model into a small one. The selection experiment carried each route on at
least two candidates, so that a route failing could be told apart from a candidate failing.

\newcommand{\objbox}[1]{\parbox[c][21mm][c]{27mm}{\centering\scriptsize #1}}
\newcommand{\fambox}[1]{\parbox[c][12mm][c]{27mm}{\centering\scriptsize\bfseries #1}}
\begin{figure}[H]
\centering
\resizebox{\textwidth}{!}{%
\begin{tikzpicture}[
  box/.style={draw=cGrey!70, fill=white, rounded corners=2pt, inner sep=0pt},
  hdr/.style={draw=cTeal!80, fill=cTeal!10, rounded corners=2pt, inner sep=0pt},
]
\node[hdr] (root) at (0,0) {\fambox{ways to learn\\without labels}};
\node[hdr] (inv) at (-6.0,-2.0) {\fambox{view-invariance\\(texture-averse)}};
\node[hdr] (mim) at ( 1.5,-2.0) {\fambox{masked\\reconstruction\\(texture-hungry)}};
\node[hdr] (reg) at ( 7.5,-2.0) {\fambox{statistical\\regularisation}};
\draw[flow] (root) -- (inv); \draw[flow] (root) -- (mim); \draw[flow] (root) -- (reg);
\node[box] (dino)  at (-7.5,-4.6) {\objbox{DINO \citep{caron2021dino}\\self-distillation\\\textbf{convnets: yes}\\per-location: no}};
\node[box] (v23)   at (-4.5,-4.6) {\objbox{DINOv2 \citep{oquab2024dinov2}\\patch-level loss\\\textbf{convnets: via}\\distilled students\\per-location: yes}};
\node[box] (mae)   at (-1.5,-4.6) {\objbox{MAE \citep{he2022mae}\\patch reconstruction\\transformer-only\\convnets: no}};
\node[box] (spark) at ( 1.5,-4.6) {\objbox{SparK \citep{tian2023spark}\\sparse masked conv\\\textbf{convnets: yes}\\per-location: yes}};
\node[box] (fcmae) at ( 4.5,-4.6) {\objbox{ConvNeXt-V2\\FCMAE \citep{woo2023convnextv2}\\GRN guard\\\textbf{convnets: yes}\\per-location: yes}};
\node[box] (vic)   at ( 7.5,-4.6) {\objbox{VICReg \citep{bardes2022vicreg}\\variance term, local\\\textbf{convnets: yes}\\per-location: yes}};
\draw[flow] (inv) -- (dino); \draw[flow] (inv) -- (v23);
\draw[flow] (mim) -- (mae); \draw[flow] (mim) -- (spark); \draw[flow] (mim) -- (fcmae);
\draw[flow] (reg) -- (vic);
\end{tikzpicture}}
\caption{Families of label-free training objectives, keyed to what matters here: does
the recipe work on a convolutional network, does it teach every location of the feature
map (``per-location''), and does it have a built-in guard against the features collapsing
to the same value. The masked-reconstruction family with that guard is the one woma uses.}
\label{fig:objectives}
\end{figure}
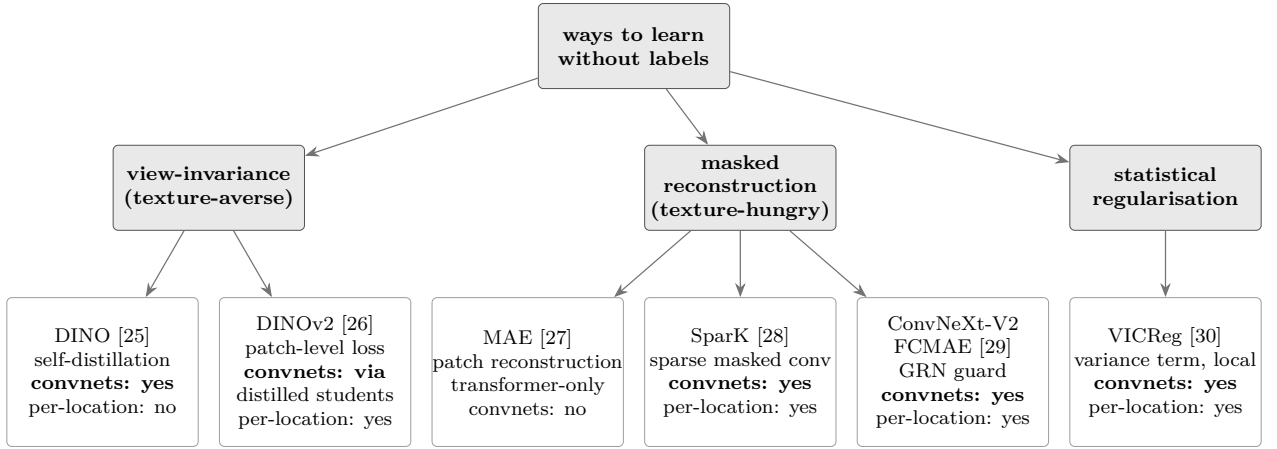
\begin{figure}[H]
\centering
\resizebox{\textwidth}{!}{%
\begin{tikzpicture}[x=1cm,y=1cm, pstep/.append style={minimum height=11mm}]
\node[pstep, text width=36mm] (a1) at (0,1.15) {random start};
\node[pstep, text width=36mm] (a2) at (0,-0.05) {long label-free
 run on the held pool};
\node[pstep, text width=36mm] (a3) at (0,-1.45) {the slowest route;\\viable at our pool size};
\draw[flow] (a1)--(a2); \draw[flow] (a2)--(a3);
\node[pstep, text width=36mm] (b1) at (5.4,1.15) {public weights\\
 (supervised or label-free)};
\node[pstep, text width=36mm] (b2) at (5.4,-0.05) {short in-domain\\
 continuation};
\node[pstep, text width=36mm] (b3) at (5.4,-1.45) {converges far faster than\\a random start \citep{reed2022hpt}};
\draw[flow] (b1)--(b2); \draw[flow] (b2)--(b3);
\node[pstep, text width=36mm] (c1) at (10.8,1.15) {foundation teacher\\(in-domain if the licence fits)};
\node[pstep, text width=36mm] (c2) at (10.8,-0.05) {distil into the
 deploy-size\\foundation model on the pool};
\node[pstep, text width=36mm] (c3) at (10.8,-1.45) {how current real-time\\detectors are built};
\draw[flow] (c1)--(c2); \draw[flow] (c2)--(c3);
\begin{scope}[on background layer]
\node[lane, fit=(a1)(a3)] (la) {};
\node[lane, fit=(b1)(b3)] (lb) {};
\node[lane, fit=(c1)(c3)] (lc) {};
\end{scope}
\node[font=\scriptsize\bfseries, text=cGrey!50!black, above=1mm of la]
 {Route A --- from scratch};
\node[font=\scriptsize\bfseries, text=cGrey!50!black, above=1mm of lb]
 {Route B --- initialise, adapt};
\node[font=\scriptsize\bfseries, text=cGrey!50!black, above=1mm of lc]
 {Route C --- distil a foundation};
\end{tikzpicture}}
\caption{Three routes to a pretrained foundation model. They compose: Route B can prepare the
teacher that Route C distils. Every route was carried by at least two of the eight
candidates, so a route failing could be told apart from a candidate failing.}
\label{fig:routes}
\end{figure}
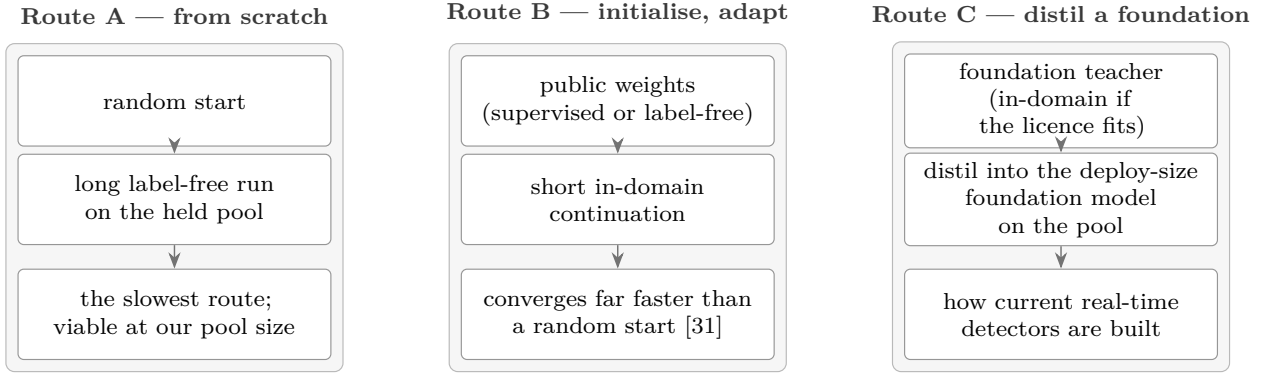

\paragraph{Design principles.} Our review compressed into eight rules, and those rules generated
our candidate set. Measure cost on target hardware, never from operation counts. Prefer public
weights or distillation for speed, keeping training from scratch as a control our pool size
allows \citep{reed2022hpt}. Pretrain a convolutional network only with an objective that teaches
every location and guards against collapse
\citep{tian2023spark,woo2023convnextv2,bardes2022vicreg}. Favour reconstruction objectives for
texture-defined lesions. Hold an unbroken additive identity path through every stage
\citep{he2016resnet}. Anti-alias each downsampling step \citep{zhang2019blurpool}, since aliasing
causes flicker and flicker causes false alarms. Give a pooled endpoint real width. And keep our
operator set boring, so a network compiles to any standard engine.

\paragraph{Candidate set.} Table~\ref{tab:cands} and Figure~\ref{fig:cands1} list the
eight candidates. They are one pick per family of a 2025--26 landscape our survey found deployable within our
latency requirement: plain residual networks, plain networks that fold into one branch at
deployment, cross-stage-partial trunks that real-time detectors are built from, modernised
convnets, and a mobile hybrid. Hierarchical vision transformers and state-space encoders were
surveyed and excluded, transformers for unproven latency and quantisation at this resolution,
state-space encoders for needing a custom kernel our engine rule forbids. This set samples every
axis selection must price at least twice: topology, pretraining route, context at the deepest
scale (a matched triplet 4, 6 and 8 differs in exactly one thing there) and deployment regime.

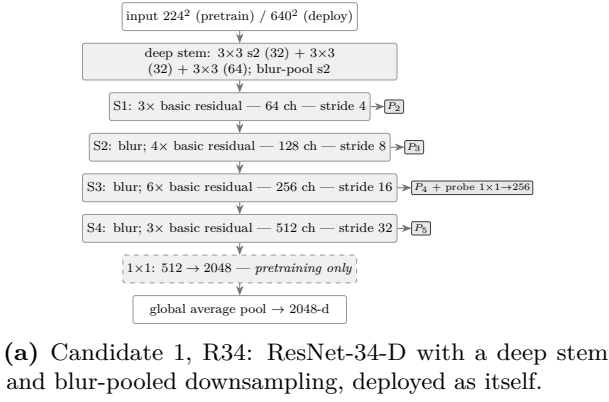
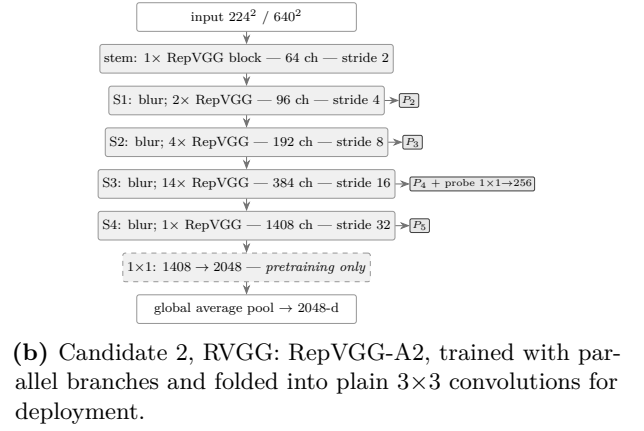
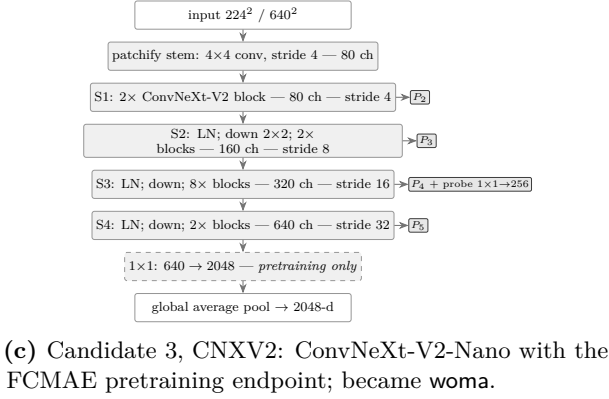
\begin{table}[t]
\centering\footnotesize
\caption{Eight candidates and a Stage-0 latency gate (batch 1, 640$^2$, f16, one forward of the network alone on an RTX 3090 as the engine proxy; the pass mark is 8\,ms for network and neck). Route: A from scratch, B public weights then adaptation, C distillation. Candidate 8's kernel path was deferred by design, so it was screened for representation quality only.}
\label{tab:cands}
\begin{tabularx}{\textwidth}{c l Y c r r >{\raggedright\arraybackslash}p{36mm}}
\toprule
\# & name & strategy & route & params & ms & gate \\
\midrule
1 & R34 & ResNet-34-D, a plain residual network deployed as itself & B & 21.8M & 2.22 & pass \\
2 & RVGG & RepVGG-A2, trained with branches, folded to a plain 3$\times$3 stack & B & 25.5M & 1.79 & pass \\
3 & CNXV2 & ConvNeXt-V2-Nano, a masked-autoencoder convnet & B & 15.6M & 3.54 & pass, flagged for its depthwise convolutions \\
4 & FMS & CSP-R-13 distilled from an in-domain foundation teacher & C & 12.7M & 1.90 & pass \\
5 & CSPR & CSP-R-22, a detection trunk made safe for self-supervision & A & 20.7M & 2.42 & pass, at the line \\
6 & CSPR-A & candidate 4's body with two attention blocks at the deepest scale & C & 19.3M & 1.97 & pass \\
7 & RVT & RepViT-M1.1, a reparameterised mobile hybrid & B & 8.2M & 4.14 & pass, the mobile-regime trade-off made visible \\
8 & CSPR-S & candidate 4's body with a state-space layer at the deepest scale & C & 14M & --- & deferred \\
\bottomrule
\end{tabularx}
\end{table}

\input{arch_figs_bw}

\paragraph{Plan, and its rules.} Figure~\ref{fig:master} shows this project whole, with its two
gates; a guiding idea is to spend measurements before GPU-hours and GPU-hours before weeks.
Stage 0 measures before anything trains and can already kill a candidate, for a latency miss in
its deployable form or training throughput too low for a twelve-hour run to mean anything. Stage 1 trains every candidate for the same wall-clock
budget and probes the frozen network at fixed intervals with two tests that need no
fine-tuning: a linear station classifier on GastroHUN \citep{bravo2025gastrohun} (macro $F_1$ over 22
stations) and a linear polyp-outline head on Kvasir-SEG \citep{jha2020kvasirseg} (Dice), each read against its untrained floor
(0.056 and the image-independent null of 0.390). A third instrument, the \emph{effective
rank} of the dense feature map, watches for collapse, with a warning line at 64. Four views of each run (equal clock; matched images seen; a route-optimal second wave; and a
short adaptability fine-tune of a whole network through one outline head, twelve epochs for every
candidate) combine into a single ranked ladder. Our advance rule was fixed in advance: top two
rungs go on, provided each clears three times its random-network floor, ends above that
effective-rank warning line, and beats twice chance on our station probe. Stage 2 gives both
finalists 72 hours each plus a short supervised fine-tune; whichever wins on detection and station
together goes on, ties broken by measured latency. Winner's route then runs to its stopping
rule on the full pool: every twenty minutes we probe that frozen network as above, and a run
stops once both probes have gained less than 0.005 over their last 3{,}000 steps, which is less than the spread between two random seeds.

\subsection{Fine-tuned models}
\label{sec:assistants}

One forward of the foundation model serves every head (Table~\ref{tab:heads}). Detection
uses a light feature-pyramid neck \citep{lin2017fpn} under an anchor-free head of the
YOLOv8 kind \citep{jocher2023yolov8}, with a single class in both models: a two-class
upper-GI variant (Barrett's segment against a focal lesion) lost, because a focal lesion
often sits inside a Barrett's field and one detector cannot serve a field and a focus at
once. Polyp type and findings are read per frame in this version. Figures~\ref{fig:wfcolon}
and \ref{fig:wfupper} give each model's complete recipe: training sets,
optimiser and schedule, augmentation, what is measured while a run trains, what is measured
on held-out data for every snapshot, and the decision rule. Three choices deserve a sentence.
Each training step draws one task by weight and runs one batch of it, so our foundation model
sees every head's gradient in proportion to those weights and no head's batch statistics leak
into another's. The foundation model is pretrained while the heads start
from random weights, so heads learn at ten times its rate. Weights we evaluate and save are a running average of live ones. Each model's shipped snapshot is chosen
on its held-out record: lesion gate filters first, then most pass marks met wins. Training
curves are never consulted.

Every arm reported here was fine-tuned on one machine at one setting: three cards, four images
each, effective batch twelve. The first pass was not --- candidates 1 and 7 ran on the second box
while the first was busy, and two cards reach twelve only at six images each. Normalisation
statistics inside a trunk are estimated per replica and never synchronised, so a replica sees four
samples or six, and for a trunk built of convolution and batch normalisation that is no detail.
Both arms were re-run at four and it is the re-runs that Table~\ref{tab:arms} reports; the earlier
pair is kept, and what it now measures is the split. Candidate 1, a residual network with batch
normalisation throughout, reads PolypGen 0.860--0.889 at six across all ten snapshots and
0.912--0.970 at four: two bands that do not meet at any epoch, one wholly under the pass mark and
one wholly over it, while every other instrument moves by less than 0.03. Candidate 7, a
transformer with layer normalisation, does not move --- 0.953--0.974 against 0.944--0.973. Machine
and split changed together, on the same model of card and byte-identical data, and a pair cannot
separate them; the matched re-runs are the answer to that, not an argument that it does not
matter.

Repeats of one setting settle a different question: which instruments tolerate a single run at all.
Across two seeds PolypGen sensitivity moves by at most 0.015 and in-domain detection by 0.023,
against 0.18 for held-out segment recall --- as large as the 0.19 that separates candidate 2's two
batch settings, so we do not read that instrument as ordered between single arms. Detection
comparisons in this paper therefore rest on ground the seeds support; segment-recall differences
smaller than about 0.2 do not, and we read them as indistinguishable rather than ordered.

\subsection{Evaluation}
\label{sec:evalprotocol}

\paragraph{Held-out instruments and metrics.} Every result is read on data training never saw:
for our colon model PolypGen \citep{ali2023polypgen}, REAL-Colon study 4
\citep{biffi2024realcolon} (15 full videos), 15 held-out CAS-Colon videos
\citep{song2025cascolon}, HyperKvasir's landmark folders \citep{borgli2020hyperkvasir}, KUMC's
held-out patients \citep{li2021kumc} and three other-hospital PraNet outline sets; for our upper-GI
model GastroHUN's validation and test patients, 39 EDD2020 \citep{ali2020edd2020} neoplasia frames
against 1{,}347 clean frames from another hospital, UGIAD's published test split
\citep{chan2025fastugi} and half of GastroVision \citep{jha2023gastrovision}.
\emph{Sensitivity} is the share of true lesions found, \emph{specificity} the share of clean frames
left alone, \emph{macro recall} that share averaged over classes, and \emph{Dice} the overlap of a
predicted outline with a true one. Every candidate carried to fine-tuning ran both recipes
unchanged and was scored on these instruments at every second epoch.

\paragraph{Speed protocol.} All four engines were timed in one sitting on one machine, on a card at
its factory 370\,W limit with no display attached; nothing else ran, and both cards were idle before
and after. Foundation-model rows ran PyTorch first and frame-to-results rows numbat first, so
neither system always met a cold box. Each engine runs 300 synchronised iterations after 30
warm-ups on the same layers: the same ConvNeXt-V2-Nano from \code{timm} for the foundation-model
rows, and for the fine-tuned rows our own layers end to end --- foundation model, neck, every head
with its class counts, the box decode and non-maximum suppression (NMS) --- from a frame in host
memory to results in host memory. Every cell was measured three times and we report a median of
three medians.

What those repeats show matters more than their number. Each numbat repeat is a separate launch,
starting its allocator and selecting its kernels afresh, while framework repeats run inside one
process after their warm-ups --- friendlier of two arrangements, and not ours. Across all 31
timed cells, widest spread between repeats is 0.067\,ms and median spread 0.014\,ms, and
in all 31 engine-to-engine comparisons numbat's \emph{slowest} repeat is faster than the
\emph{fastest} repeat of the engine it is set against: no pair of distributions overlaps anywhere in
the table. The closest case is the bf16 foundation model, where numbat's worst reading of
2.49\,ms still beats TensorRT's best of 2.57.

PyTorch is version 2.14.0 with CUDA 13 in eager mode, channels-last for its 16-bit rows; ONNX
Runtime 1.29.0 and TensorRT 11.3.0 read one exported graph of those layers, TensorRT as an engine
built at optimisation level 5; every engine's outputs were checked against PyTorch's on a fixed
frame before anything was timed. numbat is the 0.9.14 release build, timed as the program's own
stages with a device sync after each stage, over 151-frame clips.

\section{Results}
\label{sec:results}

\subsection{Screening outcome}
\label{sec:screening}

Table~\ref{tab:screen} gives our Stage-1 field. Candidate 3 ranked first: highest frozen
station probe (0.772), an outline probe within 0.01 of best, effective rank well clear of its
warning line, and highest fine-tuned Dice (0.869) in that adaptability
test. Candidate 2 ranked second. Candidate 5, the only network trained from a random start,
sat fifth on the frozen probes but showed the largest movement under fine-tuning
($+$0.268), which is what a detection-style trunk trained from scratch is expected to do,
and it was carried forward as the from-scratch control. Three distillation candidates (4, 6 and 8) went unranked: their features collapsed to an effective rank near 6, and the
cause was traced to the teacher's dense target itself, which at matched positions carries
an effective rank of 20.7, three points above what an untrained student already has. A
whitening term repaired the collapse (effective rank 66 at step 1{,}000), but too late for
the ladder. Candidate 3 then won the confirmation on both detection and station, and its
route (public pretrain-only weights of ConvNeXt-V2-Nano, self-supervised on ImageNet
photographs \citep{wightman2019timm,russakovsky2015imagenet}, continued with the same masked
autoencoder on the one-million-frame GastroNet-5M holding \citep{boers2024gastronet} plus
in-house video) stopped under its rule at step 11{,}500. That checkpoint is \woma. Its shape
is that of ConvNeXt-V2-Nano (Figure~\ref{fig:cand3}): a stem that turns each 4$\times$4
patch into 80 features, four stages of 2, 2, 8 and 2 blocks at 80, 160, 320 and 640
channels, 15.6 million parameters; each block is a 7$\times$7 depthwise convolution, a
normalisation over channels, two 1$\times$1 convolutions with a GELU (Gaussian error linear
unit) between them, and the global response normalisation (GRN) that keeps channels
diverse. Heads read stage outputs at strides 8, 16 and 32, called $P_3$ to $P_5$.

\begin{table}[t]
\centering\footnotesize
\caption{Stage-1 screening, all eight candidates at the exit of their equal-clock run: the two frozen probes (station macro $F_1$ on GastroHUN, linear Dice on Kvasir-SEG), the effective rank of the dense map, the Dice after the twelve-epoch adaptability fine-tune with an identical head, and the outcome. Floors: station 0.056, Dice 0.390 (image-independent null), effective-rank warning line 64.}
\label{tab:screen}
\begin{tabularx}{\textwidth}{l l r r r r Y}
\toprule
candidate & route & station & Dice & eff.\ rank & fine-tuned Dice & outcome \\
\midrule
3 CNXV2 & B, masked autoencoder & \textbf{0.772} & 0.731 & 101.6 & \textbf{0.869} & rung 1; won the confirmation; became \woma \\
2 RVGG & B, ImageNet weights & 0.761 & 0.717 & 105.3 & 0.862 & rung 2; finalist \\
1 R34 & B, ImageNet weights & 0.698 & \textbf{0.739} & 114.8 & 0.853 & ranked third \\
7 RVT & B, ImageNet weights & 0.713 & 0.735 & 107.5 & 0.833 & ranked fourth \\
5 CSPR & A, from scratch & 0.735 & 0.524 & 77.4 & 0.792 ($+$0.268) & fifth; carried as the from-scratch control and fine-tuned on both models \\
4 FMS & C, distilled & 0.459 & 0.081 & 6.2 & --- & unranked: collapsed on the teacher's target \\
6 CSPR-A & C, distilled & 0.517 & 0.204 & --- & --- & unranked: collapsed \\
8 CSPR-S & C, distilled & 0.240 & 0.053 & --- & --- & unranked: collapsed \\
\bottomrule
\end{tabularx}
\end{table}

\subsection{Other candidates, same recipes}
\label{sec:runnerup}

A screening is a claim about what a trunk is worth downstream, and the only way to test it is to
spend the fine-tunes. Table~\ref{tab:arms} puts every candidate through both product recipes on the
same data, schedule, augmentation, evaluation, pass marks and machine, four images per replica
throughout (Section~\ref{sec:assistants}); Figures~\ref{fig:armcolon} and \ref{fig:armupper} plot
each arm per snapshot on every held-out instrument.

\emph{Finding lesions barely separates them.} Every candidate's gastroscopy arm clears the
lesion gate, several above \woma{} (candidate 1 at 0.974, candidate 2 at 1.000, against 0.949),
and every colonoscopy arm clears its detection gate at every snapshot but one --- \woma's own
epoch 20, at 0.898, where candidate 5 still reads 0.934, having matched \woma{} at the peak
(0.969 against 0.970) and faded less. A detector head on a reasonable trunk finds lesions.

\emph{Describing a frame separates them completely, and screening predicted that order.}
Stage 1's frozen station probe ranked all five candidates
3 $>$ 2 $>$ 5 $>$ 7 $>$ 1 (0.772, 0.761, 0.735, 0.713, 0.698, Table~\ref{tab:screen}).
Their landmark-region accuracies after a full gastroscopy fine-tune fall in the same order,
with no inversion: 0.923, 0.738, 0.662, 0.634, 0.575. A twelve-hour probe on frozen features
ordered five trunks exactly as two weeks of fine-tuning did, which is the case for spending
measurements before GPU-hours. \woma{} is also the only candidate whose region accuracy
clears its bar at all, by a margin no other comes within 0.15 of.

\emph{One class can hide that.} Candidate 5's caecum recall on another centre climbs from 0.46 to
0.95 through training while its retroflexed-rectum recall falls from 0.42 to 0.11 and its
ten-segment macro declines from 0.43 to 0.35: it answers caecum more often rather than recognising
more segments, where \woma's macro holds between 0.48 and 0.52. Read on caecum alone it would
look like a better station model, which is why we fixed that pass mark on macro recall before any
run.

\emph{Candidate 1 reads that difference most clearly}, being no from-scratch control but a
supervised ImageNet residual network, most conventional trunk in our set.
Its gastroscopy arm flags lesions as well as anything here --- lesion gate 0.974, above
\woma{} --- yet names landmark region on 57.5\,\% of frames where \woma{} names it on
92.3\,\%, reads 23-code macro at 0.415 against 0.822, and meets two of six bars against five.
Its colonoscopy arm is the counter-case and we report it as one: at the matched setting it clears
the detection gate at every snapshot, meets the same two bars \woma{} meets, and is not separated
from \woma{} by more than 0.08 on any of the four colonoscopy instruments. An earlier run of this
arm at six images per replica missed the gate at every snapshot; that was the split, not the trunk
(Section~\ref{sec:assistants}). What supervised ImageNet features do not substitute for is
therefore narrower than a whole trunk's worth, and sharper for being narrow: not finding a lesion,
but naming where the frame was taken and what else is in it.

\emph{Candidate 2 clears the gate and still loses on everything the gate does not test.} Its
colonoscopy arm meets both detection marks at epoch 2 --- 0.943 of PolypGen polyps found at
precision $\geq$0.85, and 0.947 of REAL-Colon polyps at two false alarms per procedure --- and
meets two bars, the count \woma{} meets. Every descriptive head reads lower: other-hospital outline
Dice 0.616 against 0.807, ten-segment recall 0.436 against 0.521, polyp type 0.583 against 0.634.
Nor does it hold. By epoch 20 segment recall is 0.101 and Dice 0.305 while detection still reads
0.930, so the rule selects its second snapshot because that is the last one worth having. A trunk
can keep what a detector needs while shedding what every other head reads, and this is the arm that
shows it plainly.

The comparison also exposed a defect in our trainer that only a network with batch
normalisation can show; it was fixed and verified before the candidate-5 runs were repeated,
and no \woma{} number changed.

\begin{table}[t]
\centering\footnotesize
\caption{Every candidate carried through both product recipes, identical in data, schedule, augmentation, evaluation and pass marks; only the foundation model differs. Each row is the snapshot the pre-registered rule selects --- pass the anatomy's gate, then meet the most bars. $\star$ marks a bar met; \emph{sel.} is the snapshot, \emph{met} counts the gated bars. A dash is an instrument not yet scored for that snapshot.}
\label{tab:arms}
\setlength{\tabcolsep}{4pt}
\begin{tabular}{l c c c c c c c c}
\toprule
colonoscopy arm & sel. & PolypGen & REAL-Colon & Dice & segment & type & prep & met \\
\midrule
\woma{} (3 CNXV2) & 6 & 0.960$^{\star}$ & 1.000$^{\star}$ & 0.807 & 0.521 & 0.634 & 0.737 & 2 \\
5 CSPR & 8 & 0.959$^{\star}$ & 1.000$^{\star}$ & 0.732 & 0.397 & 0.527 & 0.561 & 2 \\
1 R34 & 8 & 0.954$^{\star}$ & 1.000$^{\star}$ & 0.749 & 0.449 & 0.629 & 0.666 & 2 \\
2 RVGG & 2 & 0.943$^{\star}$ & 0.947$^{\star}$ & 0.616 & 0.436 & 0.583 & 0.715 & 2 \\
7 RVT & 2 & 0.955$^{\star}$ & 0.947$^{\star}$ & 0.708 & 0.446 & 0.671 & 0.702 & 2 \\
\midrule
gastroscopy arm & sel. & gate & region & 23-code & findings & UGIAD & GastroV. & met \\
\midrule
\woma{} (3 CNXV2) & \texttt{asm10c2} & 0.949$^{\star}$ & 0.923$^{\star}$ & 0.822 & 0.877$^{\star}$ & 0.932$^{\star}$ & 0.861$^{\star}$ & 5 \\
5 CSPR & \texttt{asm14c4} & 0.974$^{\star}$ & 0.662 & 0.481 & 0.755 & 0.693 & 0.786 & 1 \\
1 R34 & \texttt{asm18c2} & 0.974$^{\star}$ & 0.575 & 0.415 & 0.837 & 0.904$^{\star}$ & 0.837 & 2 \\
2 RVGG & \texttt{asm12c4} & 0.949$^{\star}$ & 0.738 & 0.624 & 0.830 & 0.856 & 0.835 & 1 \\
7 RVT & \texttt{asm2c4} & 0.974$^{\star}$ & 0.634 & 0.359 & 0.770 & 0.741 & 0.811 & 1 \\
\midrule
\multicolumn{9}{@{}p{0.97\textwidth}@{}}{\scriptsize Colonoscopy bars: PolypGen sensitivity $\geq$0.90 at precision $\geq$0.85; REAL-Colon per-polyp sensitivity $\geq$0.93 at $\leq$2 false alarms per procedure; other-hospital Dice $\geq$0.82; segment recall $\geq$0.70. Gastroscopy bars: lesion gate sensitivity $\geq$0.93 at specificity $\geq$0.90; landmark-region accuracy $\geq$0.92; 23-code macro $\geq$0.85; findings macro $\geq$0.85; UGIAD coarse-5 $\geq$0.90; GastroVision-half $\geq$0.85.} \\
\bottomrule
\end{tabular}
\end{table}

\input{arm_curves}
\FloatBarrier

\subsection{Held-out results of the two models}
\label{sec:heldout}

Tables~\ref{tab:rescolon} and \ref{tab:resupper} give every held-out number against its
pass mark. In both models the run's own validation kept improving through the final
learning-rate decay while every other-hospital number fell: in the colon run, PolypGen
sensitivity peaked at epoch 4 (0.970) and fell to 0.898 by epoch 20 while the in-domain
detection score plateaued, and an internal composite selector would have picked epoch 9.
Epoch 6 is the colon keep, the only snapshot that passes both detection marks with margin.
Upper-GI heads reach their marks at different epochs, so what ships is an assembly: epoch 10's foundation model, detector, outline and station heads with a findings
head retrained for two epochs on that frozen network; heads retrained the same way on the
epoch-4, 6 and 8 networks clear every mark too.

\begin{table}[htbp]
\centering\footnotesize
\caption{Colon model, epoch 6 as shipped, against its pass marks, every number read on data
the training never saw. 95\,\% confidence intervals are Wilson score intervals from the frame or
polyp counts \citep{wilson1927}.}
\label{tab:rescolon}
\setlength{\tabcolsep}{5pt}
\begin{tabular}{p{7.1cm} r c r c}
\toprule
instrument & value & 95\,\% CI & mark & met \\
\midrule
\multicolumn{5}{@{}l}{\emph{polyp detection}} \\
PolypGen sensitivity, precision $\geq$0.85 & 0.960 & 0.948--0.969 & $\geq$0.90 & yes \\
PolypGen sensitivity, half the clean frames flagged & 0.897 & --- & --- & reported \\
REAL-Colon per-polyp, 1.6 alarms/procedure & 1.000 & 0.83--1.00 & $\geq$0.93 & yes \\
REAL-Colon per-polyp, 0.2 alarms/procedure & 17 of 19 & --- & --- & reported \\
\midrule
\multicolumn{5}{@{}l}{\emph{polyp outline, Dice}} \\
Kvasir-SEG, in domain & 0.907 & --- & --- & reported \\
CVC-ClinicDB \citep{bernal2015clinicdb}, in domain & 0.900 & --- & --- & reported \\
CVC-ColonDB, another hospital & 0.748 & --- & --- & reported \\
ETIS, another hospital & 0.752 & --- & --- & reported \\
CVC-300, another hospital & 0.922 & --- & --- & reported \\
mean of the three other-hospital sets & 0.807 & --- & $\geq$0.82 & no, by 0.013 \\
\midrule
\multicolumn{5}{@{}l}{\emph{colon segment, 15 held-out videos}} \\
ten segments, macro recall & 0.521 & --- & $\geq$0.70 & no \\
ten segments, with an 8-second memory & 0.574 & --- & --- & reported \\
six merged classes, accuracy (Saito protocol) & 0.655 & --- & 0.666 & parity \\
HyperKvasir, another hospital: caecum recall & 0.737 & --- & --- & reported \\
HyperKvasir: retroflexed-rectum recall & 0.529 & --- & --- & reported \\
HyperKvasir: terminal-ileum recall (9 frames) & 0.220 & --- & --- & reported \\
\midrule
\multicolumn{5}{@{}l}{\emph{polyp type}} \\
per lesion, REAL-Colon: adenoma & 4 of 15 & --- & --- & no \\
per lesion, REAL-Colon: hyperplastic & 2 of 2 & --- & --- & --- \\
per lesion, REAL-Colon: serrated & 0 of 2 & --- & --- & --- \\
per frame, REAL-Colon: macro recall & 0.617 & --- & $\geq$0.80 & no \\
per frame, KUMC held-out patients: macro recall & 0.635 & --- & $\geq$0.80 & no \\
\midrule
\multicolumn{5}{@{}l}{\emph{bowel preparation}} \\
macro recall over the four Boston classes & 0.737 & --- & --- & reported \\
weighted kappa & 0.845 & --- & --- & reported \\
grouped accuracy, adequate against not & 0.985 & --- & --- & reported \\
\midrule
\multicolumn{5}{@{}p{0.97\textwidth}@{}}{\scriptsize PolypGen is 1{,}347 polyp frames and 193
clean frames from six hospitals: at the operating point above, 1{,}293 polyps are found and 54
missed, of which only 20 carry no overlapping box at all (Section~\ref{sec:discussion}). REAL-Colon
study 4 is 15 full videos holding 19 polyps, scored per polyp with a two-second persistence rule;
the mark allows two sustained false alarms per procedure and the shipped model runs at 1.6.} \\
\bottomrule
\end{tabular}
\end{table}

\begin{table}[htbp]
\centering\footnotesize
\caption{Upper-GI model, assembled epoch-10 snapshot as shipped, against its pass marks, on
data the training never saw. 95\,\% CI as in Table~\ref{tab:rescolon}. The tiers are the order the
selection rule reads the marks in: a snapshot must clear the lesion gate first.}
\label{tab:resupper}
\setlength{\tabcolsep}{5pt}
\begin{tabular}{p{7.1cm} r c r c}
\toprule
instrument & value & 95\,\% CI & mark & met \\
\midrule
\multicolumn{5}{@{}l}{\emph{station, unseen patients --- tier 1}} \\
validation, 793 frames: landmark-region accuracy & 0.923 & 0.902--0.940 & $\geq$0.92 & yes \\
validation: landmark-region macro recall & 0.896 & --- & --- & reported \\
validation: 23-code macro recall & 0.822 & --- & $\geq$0.85 & no \\
validation: 23-code accuracy & 0.836 & --- & --- & reported \\
test, 803 frames from 59 other patients, read once & 0.917 & 0.895--0.934 & $\geq$0.92 & yes \\
test: 23-code macro recall & 0.834 & --- & --- & reported \\
\midrule
\multicolumn{5}{@{}l}{\emph{lesion gate and findings --- tier 2}} \\
sensitivity, 39 neoplasia frames & 0.949 & 0.83--0.99 & $\geq$0.93 & yes \\
specificity, 1{,}347 clean frames, another hospital & 0.912 & 0.895--0.926 & $\geq$0.90 & yes \\
findings, validation macro recall over seven classes & 0.877 & --- & $\geq$0.85 & yes \\
findings, UGIAD test: coarse-5 accuracy & 0.932 & 0.910--0.950 & $\geq$0.90 & yes \\
findings, UGIAD test: fine accuracy & 0.866 & --- & --- & reported \\
findings, UGIAD test: fine macro recall & 0.828 & --- & --- & reported \\
\midrule
\multicolumn{5}{@{}l}{\emph{other-hospital findings --- tier 3}} \\
GastroVision half, two other hospitals: accuracy & 0.861 & 0.840--0.880 & $\geq$0.85 & yes \\
\midrule
\multicolumn{5}{@{}l}{\emph{outline}} \\
EDD2020 validation, 52 masks: Dice & 0.769 & --- & --- & reported \\
\midrule
\multicolumn{5}{@{}p{0.97\textwidth}@{}}{\scriptsize The station corpus is GastroHUN's patient
split \citep{bravo2025gastrohun}; the landmark region merges the 23 codes into the seven anatomical
regions a report names. The lesion gate is EDD2020's neoplasia frames \citep{ali2020edd2020} against
lesion-free landmark frames from HyperKvasir \citep{borgli2020hyperkvasir}, a hit counted at a quarter
overlap; 37 of the 39 are flagged. UGIAD is 606 frames of its published test split
\citep{chan2025fastugi} and GastroVision half is 1{,}152 frames \citep{jha2023gastrovision}. The
outline head is research-use-only because EDD2020 is.} \\
\bottomrule
\end{tabular}
\end{table}

Both detection marks are met. Read that per-frame number with its operating point: at the loosest threshold that holds precision at 0.85, nearly all of PolypGen's 193 clean frames receive some box,
which is why per-procedure counting is what a clinician should weigh and why our program ships at
stricter confidence. Outlines miss their other-hospital mark by 0.013, on the two sets with the smallest and flattest polyps. Segment recall reads parity with published single-frame results on a
merged six-class protocol, and 0.52 across ten classes; its errors sit in adjacent tube segments
one frame cannot separate, and an 8-second memory adds 0.05. Polyp type falls far short on a class
that matters, trained on 256 serrated frames. Our upper-GI model meets five of five gated marks in
one set of weights, and region accuracy holds up on separate test patients; its 23-code macro is
reported against its line, since codes it misses are neighbouring views of one landmark that
annotators themselves disagreed on.

\subsection{Speed}
\label{sec:speed}

\begin{table}[t]
\centering\footnotesize
\caption{numbat against the engines a deployment is measured against, on the same GPU, the same layers, in one sitting under the protocol of Section~\ref{sec:evalprotocol}. Each entry is the median of three repeats with their standard deviation, so every number in this table is a distribution rather than one run. The foundation model is one ConvNeXt-V2-Nano forward at 1$\times$3$\times$640$\times$640 with the input already on the card, in milliseconds; each fine-tuned model is a frame in host memory to boxes, mask, station and findings in host memory, in frames per second. Best in each row in bold.}
\label{tab:torch}
\setlength{\tabcolsep}{3pt}
\newcolumntype{S}{>{\centering\arraybackslash}p{2.85cm}}
\begin{tabular}{S S S S S}
\toprule
regime & numbat & PyTorch & ONNX Runtime & TensorRT \\
\midrule
\multicolumn{5}{@{}p{0.97\textwidth}@{}}{\emph{foundation model, milliseconds, lower is better}} \\
strict f32 & \textbf{4.30 $\pm$ 0.01} & 9.86 $\pm$ 0.03 & 11.94 $\pm$ 0.04 & 4.83 $\pm$ 0.01 \\
f32 with TF32 & \textbf{3.74 $\pm$ 0.01} & 6.21 $\pm$ 0.00 & 9.87 $\pm$ 0.00 & 4.46 $\pm$ 0.00 \\
bf16 & \textbf{2.49 $\pm$ 0.00} & 3.88 $\pm$ 0.01 & --- & 2.57 $\pm$ 0.00 \\
f16 & \textbf{2.42 $\pm$ 0.01} & 3.82 $\pm$ 0.02 & 7.15 $\pm$ 0.01 & 3.08 $\pm$ 0.00 \\
\midrule
\multicolumn{5}{@{}p{0.97\textwidth}@{}}{\emph{upper-GI model, frames per second}} \\
strict f32 & \textbf{161.3 $\pm$ 0.4} & 68.5 $\pm$ 0.0 & 61.7 $\pm$ 0.1 & 128.9 $\pm$ 0.1 \\
f32 with TF32 & \textbf{175.4 $\pm$ 0.5} & 96.9 $\pm$ 0.1 & 75.4 $\pm$ 0.1 & 135.6 $\pm$ 0.1 \\
bf16 & \textbf{237.0 $\pm$ 0.9} & --- & --- & 222.7 $\pm$ 1.4 \\
f16 & \textbf{243.3 $\pm$ 0.0} & 155.3 $\pm$ 0.3 & 100.2 $\pm$ 0.1 & 213.2 $\pm$ 0.9 \\
\midrule
\multicolumn{5}{@{}p{0.97\textwidth}@{}}{\emph{colon model, frames per second}} \\
strict f32 & \textbf{159.7 $\pm$ 0.1} & 68.4 $\pm$ 0.0 & 61.7 $\pm$ 0.1 & 129.0 $\pm$ 0.1 \\
f32 with TF32 & \textbf{177.9 $\pm$ 0.2} & 96.9 $\pm$ 0.0 & 75.3 $\pm$ 0.1 & 135.9 $\pm$ 0.1 \\
bf16 & \textbf{243.3 $\pm$ 0.0} & --- & --- & 224.0 $\pm$ 0.3 \\
f16 & \textbf{247.5 $\pm$ 0.6} & 154.7 $\pm$ 0.2 & 99.8 $\pm$ 0.2 & 215.3 $\pm$ 0.4 \\
\midrule
\multicolumn{5}{@{}p{0.97\textwidth}@{}}{\scriptsize TF32 is the tensor-core mode every framework turns on by default for convolutions. A dash is a regime the engine has no path for: ONNX Runtime's CUDA provider does not take a bf16 graph, and PyTorch's bf16 rows are not run end to end. numbat 0.9.14; PyTorch 2.14.0 with CUDA 13 in eager mode, channels-last for its 16-bit rows, with \code{torch.compile} at \code{max-autotune} measured beside it (3.59\,ms on the foundation model in bf16, its best compiled result, against 3.57 eager); ONNX Runtime 1.29.0 on its CUDA execution provider; TensorRT 11.3.0, engines built from the same ONNX graph at optimisation level 5. Every engine's outputs were checked against PyTorch's on a fixed frame before anything was timed. Repeats were taken in alternating order, numbat's each from a separate launch of the program; in every row numbat's slowest repeat is faster than the fastest repeat of every engine it is compared with.} \\
\bottomrule
\end{tabular}
\end{table}

Table~\ref{tab:torch} sets numbat against three programs such a system would otherwise be built
on: PyTorch, ONNX Runtime and TensorRT. Each was handed identical layers and identical weights on
one card and asked to do two jobs --- one pass of our foundation model, and a whole frame from
video memory to finished results. Four \emph{precisions} appear, meaning how many bits each number
carries: ordinary 32-bit arithmetic (\code{f32}), NVIDIA's faster 32-bit mode (\code{TF32}), and
two 16-bit formats (\code{bf16}, \code{f16}). Fewer bits means quicker arithmetic and less traffic
to memory; what that costs in accuracy we measured separately (Section~\ref{sec:discussion}).

numbat is fastest in every row. One pass of our foundation model takes TensorRT, strongest of the
three, 12\,\% longer in ordinary 32-bit, 19\,\% in TF32, 3\,\% in bf16 and 27\,\% in f16, and takes
ONNX Runtime and PyTorch 1.6 to 3.0 times longer. Frame to finished results --- every head run,
detections turned into boxes, overlapping boxes dropped --- our colon model delivers 248 frames a
second in f16 where TensorRT delivers 215, and 160 against 129 in ordinary 32-bit; our gastroscopy
model 243 against 213, and 161 against 129.

Two rounds of work got there, each timing our own code step by step beside whichever program was
beating us. Round one deleted work rather than speeding it up. A convolution block used to write
its result to memory, read it back to add a bias, read it again to apply an activation, and again
for a residual and a normalisation. Doing all of that inside the pass that already holds the
numbers removes four journeys to memory and changes no answer --- outputs stayed identical to the
last digit. Round one also taught every part of the program, not our foundation model alone, to
work in 16-bit numbers from end to end.

Round two compared single operations with TensorRT's. Ours were already quicker at 1$\times$1
convolutions, commonest operation in this network. Its 7$\times$7 \emph{depthwise} convolution ---
one filter per channel, which is what gives a ConvNeXt block its wide view of a frame --- ran five
times faster than ours, and that one operation was the whole remaining gap. Rewriting it closed
that gap: a row of output now reads its strip of the frame once into the small fast memory beside
the arithmetic units, one thread produces four neighbouring outputs, and 16-bit inputs stay 16-bit
instead of being widened to 32. Two smaller changes followed. One normalisation step is folded
into the weights of the layer after it, so it costs nothing while the program runs. And the pass
that writes an activation now hands back the totals the next step needs, instead of that step
reading the whole activation a second time to work them out.

f16 is what the program runs by default. Ordinary 32-bit stays available as a switch and is what
every held-out number in this paper was read with, and the optimised build reproduces the previous
release's 32-bit output on every frame of both test clips.

\paragraph{A build that needs nothing installed.} Everything above reaches the card through
NVIDIA's libraries --- CUDA, cuBLAS, cuDNN --- which a workstation must have installed, and kept in
step with its graphics driver. numbat can run the same weights a second way. We write each GPU
operation once and compile it to SPIR-V, a portable format for GPU programs that modern drivers
accept, then send it to the card through Vulkan, which ships inside the driver itself. Nothing
from any vendor is built in. What a site installs is then \emph{two files}, one program and one
model, and on Linux that program asks the system for its C library and loader and for nothing
else. No toolkit, no cuDNN version to match, no PyTorch, no Python. Those two files run on any
card whose driver can run Vulkan, which today means NVIDIA, AMD and Intel, and will mean Apple
silicon once we have tested that route.

Portability usually costs speed. Table~\ref{tab:vk} runs one program both ways, on one card and
on the same two clips, timed exactly as Table~\ref{tab:torch} timed the four engines --- so the
two tables can be read against each other. In 32-bit the portable build is the \emph{faster} of
the two, 185 frames a second against 162 on our colon model and 180 against 161 on our
gastroscopy model, because our own code puts a 32-bit convolution on the card's matrix-multiply
units where cuDNN takes a slower route. In 16-bit it is behind by 12 and 16\,\%, 217 against 246
and 207 against 245, and cuDNN's 7$\times$7 depthwise convolution is most of that difference. It
still clears 200 frames a second, eight times what a screen shows, and it finds the same things:
identical box counts in every cell where both planes run. A department with no CUDA, or with a
card that is not an NVIDIA one, gives up nothing in what this system finds and, in 32-bit, gains
speed.

\begin{table}[htbp]
\centering\footnotesize
\caption{numbat's two GPU planes, measured under the protocol and in the units of
Table~\ref{tab:torch}: same clips, three repeats, and the same quantity --- a whole frame from
video memory to finished results, in frames per second. Both columns were timed together on
numbat 0.9.16, and this CUDA column agrees with Table~\ref{tab:torch}'s numbat column to within
1.5\,\%, which is the spread between sittings. What each plane finds is identical: 113 boxes in
32-bit and 112 in 16-bit on the colonoscopy clip, 2 on the gastroscopy clip, in every cell where
both planes run. Faster of the two in bold.}
\label{tab:vk}
\setlength{\tabcolsep}{6pt}
\begin{tabular}{l c c c c}
\toprule
& \multicolumn{2}{c}{colon model, FPS} & \multicolumn{2}{c}{upper-GI model, FPS} \\
\cmidrule(lr){2-3}\cmidrule(lr){4-5}
regime & CUDA & portable & CUDA & portable \\
\midrule
strict f32 & 162.1 & \textbf{184.8} & 161.0 & \textbf{180.2} \\
f32 with TF32 & 179.9 & --- & 178.9 & --- \\
bf16 & 243.9 & --- & 241.0 & --- \\
f16 & \textbf{245.7} & 217.4 & \textbf{244.5} & 206.6 \\
\midrule
\multicolumn{5}{@{}p{0.93\textwidth}@{}}{\scriptsize CUDA is NVIDIA's libraries (cuBLAS, cuDNN);
portable is our own kernels through Vulkan, with no vendor library linked. A dash is a regime the
portable plane has no path for: TF32 is a mode of NVIDIA's tensor cores and has no counterpart, so
its f32 row stands for both; bf16 kernels are not written for it yet and the program stops with
\code{UnsupportedType} rather than guessing. One pass of the foundation model alone, the first
block of Table~\ref{tab:torch}, is not quoted for the portable plane: it queues work and returns
before the card has run it, so a per-stage stopwatch measures the queueing, not the arithmetic.
The whole-frame figure is sound on both planes because reading detections back forces the card to
finish.} \\
\bottomrule
\end{tabular}
\end{table}

Table~\ref{tab:speed} gives our program as delivered: what ships is one file
per model (66.6\,MB, holding foundation model, every head, class names, thresholds
and a fingerprint of every source weight) and one program on that same library,
with video decoding and encoding compiled in and no framework at runtime. It runs as a
three-stage pipeline (decode and letterbox; foundation model, heads and overlay; encode)
whose numbers are identical to a serial program on every frame, and it draws nothing that
obstructs the mucosa: four corner marks per detection, contours only for outlines, a
station checklist that ticks a site only after a one-second vote, and a badge with the
delivered rate.

\section{Discussion}
\label{sec:discussion}

\paragraph{What the systematic design bought.} Fixing requirements, candidates and decision
rules before any run changed what the experiment could find. That funnel caught a whole route
failing --- three distilled candidates collapsing on a teacher target that was smooth rather than
rich --- and could say so because two candidates carried every route, keeping route failure
distinguishable from candidate failure. It also caught what training curves hide: in both models
in-domain validation kept rising through the final decay while every other-hospital number fell,
so a selector reading those curves would have shipped a worse model. Running probes, trainer,
evaluators and runtime on one library, with identical operators in training and inference, is what
lets a shipped file reproduce its acceptance record to three decimals, and what made the speed work
possible --- profiling against TensorRT said which kernel was behind, and we could rewrite it that
day. Re-scored in f16 both files move 6 of 141 upper-GI record fields and 5 of 41 colon fields, each
by at most one frame; in bf16, 27 and 12. No pass mark changes status.

\paragraph{What the models are worth.} \woma{} is a 15.6-million-parameter network serving nine
heads across two products inside a 16.7\,ms budget. Three marks go unmet: colon segment (a
single-frame ceiling, where an 8-second memory adds 0.05 and 0.7 needs a model with memory), polyp
type (4 of 15 adenomas per lesion, on 256 serrated training frames) and other-hospital outline,
short by 0.013 and 0.03 Dice under the best published Kvasir-SEG numbers \citep{li2025vmdunet} from
single-task models several times as expensive. Other candidates make our choice of foundation
model legible: on identical recipes they find lesions as well as \woma{} yet describe a frame
worse, in the order a frozen probe had already put them in. A supervised ImageNet residual
network, most conventional among them, makes that point sharpest --- it flags lesions as well as
anything here and names a landmark region on 57.5\,\% of frames where \woma{} names it on 92.3\,\%.

\paragraph{Where they fail.} Of 1{,}347 PolypGen polyp frames the colon detector misses 54 at the
half-overlap its sensitivity is read at, but only 20 carry no overlapping box at all: two thirds of
the misses are a box on the lesion that is not tight enough, which costs a number and would not
cost a polyp. All 19 REAL-Colon lesions are found while up to two sustained false activations per
procedure are allowed; tightened to 0.2, two lesions are held only by boxes under that threshold.
The upper-GI gate flags 37 of 39 held-out neoplasia frames while leaving 91\,\% of 1{,}347 clean
frames from another hospital alone; the two it misses carry no box at a quarter overlap. The
findings head fails where its training data is thinnest and in one direction: its largest single
UGIAD error is neoplasia read as gastric inflammation (23 of 153 frames), and it recovers 7 of 32
GastroVision gastric polyps. Both classes are carried in the hundreds against tens of thousands of
normal frames, and both err towards the benign reading --- the direction that matters most in a
screening tool, and what a local fine-tune is aimed at.

\paragraph{What held-out evaluation does not settle.} Every number here was read on other
hospitals, other patients or other videos, and shipped weights were chosen on that record. That is
one step, not a destination: deciding samples are small. REAL-Colon study 4 holds 19
polyps, so 19 of 19 carries a 95\,\% lower bound of 0.83 \citep{wilson1927}, and the upper-GI gate
has 39 positive frames, mostly oesophageal stills. Neither gate sees blur, bubbles, instruments or
gastric cancers in motion, no endoscopist has used the overlay during a procedure, and the GPU is a
workstation card rather than a procedure-room box. What is missing is a full-procedure gastroscopy
set with per-lesion truth, and none is public. Data from our own institutions would
serve twice --- as unlabelled pretraining frames, and as the held-out set that finally measures a
full procedure on local equipment --- and everything here was built so it enters without a change
of method: same splits by patient, same pass marks, same held-out selection.

\section{Conclusion}
\label{sec:conclusion}

We set out to build a real-time foundation model for endoscopy as one builds a product:
requirements first, every decision written down before it was taken. Out came \woma, chosen from
eight candidates by a pre-registered experiment and trained without labels to a stopping rule,
and two fine-tuned models on it that meet most of their pass marks on data training never saw and
state the rest. Every step, from screening probes to shipped file, ran on one self-contained
library, and both foundation model and fine-tuned models run faster than PyTorch, ONNX Runtime and
TensorRT on the same GPU in every precision regime. What ships is one file per model and one
program, and on a second build that program carries no vendor library at all.

\section*{Declarations}

\paragraph{Ethics.} This study used only publicly available, de-identified datasets under their
published licences, named with their roles in Figures~\ref{fig:dataflow}, \ref{fig:wfcolon} and
\ref{fig:wfupper}; the authors collected no data from patients, and no ethical approval was
required. Datasets licensed for research use only (EDD2020, UGIAD, SAGE \citep{oli2026sage},
GastroEndoNet \citep{bitto2025gastroendonet}, Kvasir v2 \citep{pogorelov2017kvasir} and
GastroVision) trained heads that are marked as research heads and are not offered for
clinical use. The models are research artefacts and not medical devices; no endoscopist has used
them during a procedure.

\paragraph{Data and code availability.} Every held-out score is a record written by one
scorer, and every table and chart regenerates from those records. Records, both model files,
our program and its benchmark scripts are available from the corresponding author. Our
library, numbat, is described in \citet{tran2026numbat} and \citet{tran2026crossstack}.

\paragraph{Funding.} This work received no grant or funding from any agency in the public,
commercial or not-for-profit sectors.

\paragraph{Declaration of competing interests.} T.T. develops numbat and is affiliated with
CloudKites AI Lab, which may commercialise the library and the models described here. L.D.
declares no competing interests.

\paragraph{CRediT author contributions.} Thang Tran: conceptualisation, methodology, software,
investigation, data curation, validation, visualisation, writing -- original draft. Lan Dang:
conceptualisation, methodology, project administration, writing -- review and editing.

\paragraph{Acknowledgement.} AI assistance was used under our direction to perform literature
search, and to draft and edit text and figure code.

\FloatBarrier
\bibliographystyle{unsrtnat}
\bibliography{woma_system}

\clearpage
\setcounter{figure}{0}\renewcommand{\thefigure}{S\arabic{figure}}
\setcounter{table}{0}\renewcommand{\thetable}{S\arabic{table}}
\section*{Supplementary material}

\paragraph{S1. Head designs.} Table~\ref{tab:heads} gives every head's design; both models share
foundation model, the neck and the head designs, and only the class lists differ.

\begin{table}[t]
\centering\footnotesize
\caption{Head designs. Both models share the foundation model, the neck and the head designs; only the class lists differ.}
\label{tab:heads}
\setlength{\tabcolsep}{5pt}
\begin{tabular}{p{4.0cm} p{2.2cm} p{8.4cm}}
\toprule
head & reads & design and output \\
\midrule
detection & $P_3$, $P_4$, $P_5$ & feature-pyramid neck under an anchor-free head; box edges as distributions over 16 bins; one class \\
outline & $P_4$, $P_5$ & a small decoder to a 640$^2$ mask \\
station & pooled $P_5$ & two-layer classifier to 10 colon segments or 22$+$1 gastric stations \\
polyp type / findings & pooled $P_5$ & linear classifier to 4 polyp types or 7 upper-GI findings \\
bowel preparation & pooled $P_5$ & linear classifier to the 4 Boston classes \\
\midrule
\multicolumn{3}{@{}p{0.97\textwidth}@{}}{\scriptsize The neck follows \citet{lin2017fpn}, the head
\citet{jocher2023yolov8} and the distributional box edges \citet{li2020gfl}. A frame that carries no
label for a head adds no loss through it, so a head is never trained on a guess.} \\
\bottomrule
\end{tabular}
\end{table}

\paragraph{S2. How the two fine-tuned models were developed.} Figures~\ref{fig:wfcolon} and
\ref{fig:wfupper} give each model's complete recipe: data and what was kept out of training,
training settings, what is measured while a run trains, what is measured on held-out data for
every snapshot, pass marks, and the rule that chose shipped weights.

\input{workflow_figs}
\afterpage{\clearpage}

\paragraph{S3. Program as delivered.} Table~\ref{tab:speed} gives delivered rate, latency and
per-stage cost on both test clips (Section~\ref{sec:speed}).

\begin{table}[H]
\centering\footnotesize
\caption{Program as delivered, on one GPU (f16 by default; medians over three runs). Delivered is the interval between finished frames, the badge's number; the stage figures are per frame with a device sync after every GPU stage.}
\label{tab:speed}
\setlength{\tabcolsep}{5pt}
\begin{tabular}{p{7.4cm} r r}
\toprule
 & gastroscopy clip & colonoscopy clip \\
 & 1350$\times$1080, 15 fps & 316$\times$256, 30 fps \\
\midrule
delivered FPS: mean & 100.3 & 118.3 \\
delivered FPS: median & 99.2 & 117.6 \\
delivered FPS: 5th percentile & 90.4 & 114.3 \\
delivered FPS: minimum & 82.7 & 109.8 \\
frames delivered at $\geq$24 FPS & 100\,\% & 100\,\% \\
frame to results, mean ms & 6.9 & 6.9 \\
frame to results, median ms & 5.4 & 5.4 \\
latency, frame read to encoded: mean ms & 19.7 & 18.5 \\
latency, frame read to encoded: median ms & 15.3 & 14.1 \\
\midrule
\multicolumn{3}{@{}p{0.97\textwidth}@{}}{\scriptsize Clips are 151 and 150 frames. Delivered is the
interval between finished frames, the badge's number; frame to results covers upload, foundation
model, heads, decode and NMS. Stage medians on the gastroscopy clip, ms per frame: letterbox and
normalise 8.9 (CPU, producer thread), upload 0.9, foundation model 2.8 (5.4 in f32), detection head
and decode 0.55, outline head 1.1, station and findings 0.1, NMS and post 0.0, overlay 0.3, encode
6.5 mean (CPU, encoder thread).} \\
\bottomrule
\end{tabular}
\end{table}

\paragraph{S4. Station label spaces and their groupings.} Table~\ref{tab:stations} lists every
station class both models carry, and the coarser space each is also scored in. Grouping never
changes a prediction: a model answers in its own space, and grouping is applied to both the answer
and the reference before counting.

\begin{table}[t]
\centering\footnotesize
\caption{Station labels as shipped, with the coarser spaces used for reporting. Colon merges follow
the published single-frame benchmark so accuracy is comparable to it; gastroscopy regions merge the
four walls photographed at one protocol level.}
\label{tab:stations}
\setlength{\tabcolsep}{5pt}
\begin{tabular}{c l l l}
\toprule
\multicolumn{4}{@{}l}{\emph{colonoscopy --- CAS-Colon, ten classes, merged to six for comparison \citep{saito2021anatomical}}} \\
\midrule
index & label & segment & merge \\
\midrule
0 & \code{termIleum} & terminal ileum & 0 \\
1 & \code{cecum} & caecum & 1 \\
2 & \code{ascend} & ascending colon & 2 \\
3 & \code{hepFlex} & hepatic flexure & 2 \\
4 & \code{transv} & transverse colon & 2 \\
5 & \code{splFlex} & splenic flexure & 3 \\
6 & \code{descend} & descending colon & 3 \\
7 & \code{sigmoid} & sigmoid colon & 3 \\
8 & \code{rectum} & rectum & 4 \\
9 & \code{anal} & anal canal & 5 \\
\midrule
\multicolumn{4}{@{}l}{\emph{gastroscopy --- GastroHUN, 22 sites plus \code{OTHERCLASS}, merged to seven regions \citep{bravo2025gastrohun}}} \\
\midrule
region & codes & protocol level & view \\
\midrule
1 & \code{A1 L1 P1 G1} & antrum & anterograde \\
2 & \code{A2 L2 P2 G2} & distal (lower) body & anterograde \\
3 & \code{A3 L3 P3 G3} & upper-middle body & anterograde \\
4 & \code{A4 L4 P4 G4} & fundus and cardia & retroflex \\
5 & \code{A5 L5 P5} & body & retroflex \\
6 & \code{A6 L6 P6} & incisura & retroflex \\
7 & \code{OTHERCLASS} & none of the 22 & --- \\
\midrule
\multicolumn{4}{@{}p{0.97\textwidth}@{}}{\scriptsize A letter names the wall and a digit names the
protocol level: \code{A} anterior wall, \code{L} lesser curvature, \code{P} posterior wall,
\code{G} greater curvature, as the dataset's own figure defines them, following the systematic
screening protocol for the stomach \citep{yao2013sss}. Greater curvature is photographed at
four levels rather than six, giving 22 sites. Each one:
\code{A1} antrum, anterior wall;
\code{L1} antrum, lesser curvature;
\code{P1} antrum, posterior wall;
\code{G1} antrum, greater curvature;
\code{A2} lower body, anterior wall;
\code{L2} lower body, lesser curvature;
\code{P2} lower body, posterior wall;
\code{G2} lower body, greater curvature;
\code{A3} middle-upper body, anterior wall;
\code{L3} middle-upper body, lesser curvature;
\code{P3} middle-upper body, posterior wall;
\code{G3} middle-upper body, greater curvature;
\code{A4} fundus and cardia, anterior wall;
\code{L4} fundus and cardia, lesser curvature;
\code{P4} fundus and cardia, posterior wall;
\code{G4} fundus and cardia, greater curvature;
\code{A5} body in retroflexion, anterior wall;
\code{L5} body in retroflexion, lesser curvature;
\code{P5} body in retroflexion, posterior wall;
\code{A6} incisura, anterior wall;
\code{L6} incisura, lesser curvature;
\code{P6} incisura, posterior wall.
Levels 1--3 are photographed anterograde and 4--6 in retroflexion. Regions here merge the walls at one
level, which is the grouping the landmark-region figure is read in; a seventh group holds
\code{OTHERCLASS}, which the dataset assigns when the intended site is not clearly visible or a
lesion is present, and which the live program never ticks. Protocol levels are named from the
dataset paper's withdrawal sequence, which describes steps rather than naming six levels; its count
of 12 anterograde and 10 retroflex photographs matches digits 1--3 and 4--6 exactly. Colon merges
drop the benchmark's seventh class, ``indistinguishable'', which CAS-Colon has no counterpart for.
Six-class merge: 0 terminal ileum, 1 caecum, 2 ascending with hepatic flexure and transverse,
3 splenic flexure with descending and sigmoid, 4 rectum, 5 anal canal.} \\
\bottomrule
\end{tabular}
\end{table}

\paragraph{S5. Lesion, polyp-type and findings label spaces.} Table~\ref{tab:lesionlabels} lists
what each detector and each descriptive head answers in. Both detectors carry one class: a box says
where, and a separate head says what.

\begin{table}[t]
\centering\footnotesize
\caption{Lesion and finding labels as shipped, with the coarser space each is also scored in.}
\label{tab:lesionlabels}
\setlength{\tabcolsep}{5pt}
\begin{tabular}{c l l l}
\toprule
\multicolumn{4}{@{}l}{\emph{colonoscopy}} \\
\midrule
index & label & meaning & held-out reporting \\
\midrule
--- & \code{polyp} & detection, one class & --- \\
\midrule
0 & \code{adenoma} & adenomatous polyp & reported \\
1 & \code{hyperplastic} & hyperplastic polyp & reported \\
2 & \code{serrated} & sessile serrated lesion & trained, not in the held-out split \\
3 & \code{other} & any other histology & trained, not in the held-out split \\
\midrule
\multicolumn{4}{@{}l}{\emph{gastroscopy}} \\
\midrule
index & label & meaning & UGIAD coarse-five \\
\midrule
--- & \code{lesion} & detection, one class & --- \\
\midrule
0 & \code{normal} & no finding & normal \\
1 & \code{eso\_mucosal} & oesophagitis or Barrett's & oesophageal mucosal change \\
2 & \code{neoplasia} & suspicious, high-grade dysplasia, cancer & gastric lesion \\
3 & \code{gastric\_polyp} & gastric polyp & gastric lesion \\
4 & \code{gastric\_inflammatory} & ulcer, inflammation, blood & gastric lesion \\
5 & \code{gastric\_metaplasia} & intestinal metaplasia or atrophy & metaplasia \\
6 & \code{varices} & oesophageal or gastric varices & varices \\
\midrule
\multicolumn{4}{@{}p{0.97\textwidth}@{}}{\scriptsize Polyp type is a partial-label task: a frame
whose polyp has no histology appears in no class list and contributes detection loss alone, so the
head is never taught a class the record does not support. Its reported figure is macro recall over
adenoma and hyperplastic, because the held-out patient split carries no serrated or other frame at
all; both classes stay in the output space, and the model does predict them on other sets. Training keeps oesophagitis and Barrett's
apart as separate classes and the shipped head merges them into \code{eso\_mucosal}, because
neither is what the detector is asked to flag and the two are not reliably separable on a single
frame. Coarse-five is the space the UGIAD comparison is read in. Both detectors carry one class
deliberately: box and description are separate answers, so a wrong description cannot suppress a
correct box.} \\
\bottomrule
\end{tabular}
\end{table}

\paragraph{S6. Bowel preparation scale.} Table~\ref{tab:bbps} gives the published segment scale the
preparation head answers in \citep{lai2009bbps}.

\begin{table}[t]
\centering\footnotesize
\caption{Boston Bowel Preparation Scale, scored per colonic segment.}
\label{tab:bbps}
\setlength{\tabcolsep}{5pt}
\begin{tabular}{c l p{10.4cm}}
\toprule
score & label & segment as the published scale defines it \\
\midrule
0 & \code{bbps0} & Unprepared segment, mucosa not seen because of solid stool that cannot be cleared. \\
1 & \code{bbps1} & Part of the mucosa seen, other parts not, because of staining, residual stool or opaque liquid. \\
2 & \code{bbps2} & Minor residual staining, small stool fragments or opaque liquid, mucosa seen well. \\
3 & \code{bbps3} & Entire mucosa seen well, no residual staining, stool fragments or opaque liquid. \\
\midrule
\multicolumn{3}{@{}p{0.97\textwidth}@{}}{\scriptsize Adequacy, the figure reported against a pass
mark, treats 2 and 3 as adequate and 0 and 1 as inadequate. Training labels are partial where the
source is: Nerthus frames carry an exact score, HyperKvasir frames carry a grouped one --- either
\{0,1\} or \{2,3\} --- and the grouped ones train through a partial-label loss that asks only for
the group to be right, rather than guessing which of the two the annotator meant.} \\
\bottomrule
\end{tabular}
\end{table}

\end{document}